# Optimized Flux Single-Crystal Growth of the Quantum Spin Liquid Candidate $NdTa_7O_{19}$ and Other Rare-Earth Heptatantalates, $ErTa_7O_{19}$ and $GdTa_7O_{19}$

Lia Šibav[1,2], Matic Lozinšek[1,2], Zvonko Jagličić[3,4], Tina Arh[1,5], Panchanana Khuntia[6,7], Andrej Zorko[1,5] and Mirela Dragomir[1,2*]

*[1]Jožef Stefan Institute, Jamova cesta 39, 1000 Ljubljana, Slovenia*
*[2]Jožef Stefan International Postgraduate School, Jamova cesta 39, 1000 Ljubljana, Slovenia*
*[3]Faculty of Civil and Geodetic Engineering, University of Ljubljana, Jamova cesta 2, 1000 Ljubljana, Slovenia*
*[4]Institute of Mathematics, Physics and Mechanics, Jadranska cesta 19, 1000 Ljubljana, Slovenia*
*[5]Faculty of Mathematics and Physics, University of Ljubljana, 1000 Ljubljana, Slovenia*
*[6]Department of Physics, Indian Institute of Technology Madras, Chennai 600036, India*
*[6]Quantum Centre of Excellence for Diamond and Emergent Materials, Indian Institute of Technology Madras, Chennai 600036, India*

## ABSTRACT

Single crystals are essential for characterizing a wide range of magnetic states, including exotic ones such as quantum spin liquids. This study reports a flux method for growing single crystals of $NdTa_7O_{19}$, the first quantum spin liquid candidate on a triangular spin lattice with dominant Ising-like spin correlations. Purple $NdTa_7O_{19}$ single crystals with hexagonal morphology were successfully grown using a $K_2Mo_3O_{10}$–$B_2O_3$ flux. With lateral sizes up to 3.5 mm and a thickness up to 2 mm, these are the largest dimensions reported to date. The chemical composition was confirmed by powder and single-crystal X-ray diffraction along with scanning electron microscopy with energy dispersive X-ray spectroscopy. Aiming for an accurate determination of the magnetic anisotropy and its effect on the magnetic properties, $NdTa_7O_{19}$ crystals were additionally analyzed by magnetic susceptibility, revealing a substantial anisotropy without long-range magnetic ordering down to 2 K. Single crystals of two novel rare-earth heptatantalates, $ErTa_7O_{19}$ and $GdTa_7O_{19}$, were also grown and their magnetic properties investigated. The magnetic anisotropy of $ErTa_7O_{19}$ closely resembles that of isostructural $NdTa_7O_{19}$, indicating a possibility of a similar exotic magnetic ground state. In contrast, $GdTa_7O_{19}$ shows paramagnetic behavior, consistent with previous results obtained for polycrystalline samples.

## 1. INTRODUCTION

Rare-earth heptatantalates, abbreviated as $RETa_7O_{19}$, with RE representing a trivalent rare-earth element, are promising materials for non-linear optics and laser technology due to their non-centrosymmetric structure as well as high chemical and thermal stability, high concentration of optically active ions, high mechanical strength, and high thermal conductivity.[1,2,3,4] Among this series, $NdTa_7O_{19}$ is regarded as the most auspicious candidate for such applications.[2,4]

Recently, $NdTa_7O_{19}$ has also been identified as a very intriguing magnetic material, more specifically, a potential triangular antiferromagnet with an Ising quantum spin liquid ground state.[5] Low-dimensional antiferromagnets realized on a perfect triangular lattice, such as $NdTa_7O_{19}$, are highly desired as a large magnetic anisotropy could explain a possible quantum spin liquid ground state.[6,7] Indication of a spin liquid ground state in $NdTa_7O_{19}$ was found via absence of magnetic Bragg peaks in neutron powder diffraction at temperatures down to 40 mK,[5] the presence of magnetic diffuse scattering, and persistent, low-temperature spin dynamics detected via muon spectroscopy, thus combining most suitable experimental techniques that can be used to detect such elusive states.[8] The splitting of crystal-electric field levels and bulk magnetic measurements on this material further suggested the presence of large magnetic anisotropy.[5] However, since these measurements were conducted on polycrystalline $NdTa_7O_{19}$, a more precise determination of magnetic anisotropy is dependent on the availability of single crystals.[9,10,11] Single crystals would also allow for more refined magnetic measurements, *e.g.*, enabling wave-vector-resolved insight into spin correlations or detection of the anticipated fractionalized excitations. Furthermore, given that materials with frustrated lattices and enhanced quantum mechanical properties tend to be particularly sensitive to defects (*e.g.*, chemical disorder, undesirable impurities, etc.),[8,11] there is a clear preference for high-quality single crystals of $NdTa_7O_{19}$ over polycrystalline samples, where such defects are minimized.

Single crystals of $NdTa_7O_{19}$ have been previously grown for optical applications by a flux method using two different fluxes, $Li_2B_4O_7$[1,3] and $K_2Mo_3O_{10}$—with a small addition of $B_2O_3$.[3,4] The literature reports showed a preference for starting from the constituent oxides $Nd_2O_3$ and $Ta_2O_5$[1,2,3,4] instead of polycrystalline $NdTa_7O_{19}$ which could be due to the fact that solid-state synthesis resulted in non-stoichiometric powders.[4] The largest crystals grown from the oxides were limited to 1.5 mm in lateral size.[12] Starting with constituent oxides is, however, less advantageous over starting from a polycrystalline powder of the correct composition. Different dissolution rates of the constituent oxides affect concentrations of reagents in the melts and drive the melt into a different supersaturated regime from which nucleation of undesired phases takes place.[13] Indeed, the phase diagrams showed a narrow primary crystallization region of the $NdTa_7O_{19}$ phase as well as multiple regions of formation of secondary phases which included $Nd_{0.33}TaO_3$ and $NdTaO_4$.[1,3,4] This could also be the reason why these preliminary reports[1,2,3,4] only provide rough guidelines rather than exact protocols required for the growth of high-quality, millimeter-sized single crystals.

With the aim of growing $NdTa_7O_{19}$ crystals of appropriate size for magnetic measurements and other techniques that require larger crystals, such as neutron scattering or muon spectroscopy, and establishing a more precise protocol that could be extended to other rare-earth members of the $RETa_7O_{19}$ family, in this study, we have chosen to further explore a flux growth method using $K_2Mo_3O_{10}$–$B_2O_3$ as a flux.[14] This flux has a high dissolution ability due to the high chemical activity of alkali polymolybdates, as well as a relatively low melting temperature.[4] The presence of small amounts of $B_2O_3$ increases the solubility, decreases the saturation temperature and reduces the flux volatility without significantly altering the flux viscosity.[15] These benefits have previously been successfully exploited for single-crystal growth of other systems such as rare-earth borates $REAl_3(BO_3)_4$[16,17,18,19,20,21,22] or rare-earth phosphates.[23]

In this study, nearly single-phase polycrystalline $NdTa_7O_{19}$ was successfully prepared. A previously reported solid-state method[24] was optimized, resulting in a 98(1) wt% of the main phase, which was used as a starting material for flux growth of $NdTa_7O_{19}$ single crystals of predominantly hexagonal, plate-like morphologies with lateral sizes up to 3.5 mm and thicknesses up to 2 mm—the largest $NdTa_7O_{19}$ single crystals reported to date. The present flux method was further employed to grow single crystals of two other novel rare-earth heptatantalates, $ErTa_7O_{19}$ and $GdTa_7O_{19}$. The composition and morphology of all newly-grown crystals were characterized by powder X- ray diffraction (PXRD) and single-crystal X-ray diffraction (SCXRD) as well as scanning electron microscopy (SEM) coupled with energy dispersive X-ray spectroscopy (EDS). Furthermore, magnetic susceptibility measurements were employed to investigate the magnetic anisotropy.

## 2. EXPERIMENTAL

### 2.1 Solid-state synthesis

Three polycrystalline $RETa_7O_{19}$ members, $NdTa_7O_{19}$, $ErTa_7O_{19}$ and $GdTa_7O_{19}$, were synthesized by a solid-state reaction similar to a previously reported procedure.[24] Stoichiometric amounts of $Nd_2O_3$ (Thermo Scientific, 99.99%), $Er_2O_3$ (Thermo Scientific, 99.99%), and $Gd_2O_3$ (Thermo Scientific, 99.99%), together with $Ta_2O_5$ (Alpha Aesar, 99.85 or 99.99%), were weighed, hand-homogenized in an agate mortar, and pressed into pellets with a 10 mm diameter using a force of 50 kN. Prior to the synthesis, all three rare-earth oxides were pre-annealed at 1000 °C for 24 hours. Multiple synthesis cycles (5–9) were performed at temperatures $T = 900-1200$ °C with intermediate regrinding, resulting in about 8.5 g (5 mmol) of polycrystalline product. As it is not commercially available, the main flux component, the potassium molybdate $K_2Mo_3O_{10}$, (about 30 mmol) was synthesized using $K_2CO_3$ (Chempur, 99.9%) and $MoO_3$ (Merck, 99.9%) by a solid-state reaction at 500 °C, for two cycles of 72 hours each, and intermediate regrinding (Equation 1):

$$K_2CO_3 + 3\ MoO_3 \rightarrow K_2Mo_3O_{10} + CO_2 \uparrow \qquad (1)$$

### 2.2 Single-crystal growth

Polycrystalline $RETa_7O_{19}$ samples, obtained by solid-state reactions, were used as starting materials for flux-growth experiments, employing a $K_2Mo_3O_{10}$–$B_2O_3$ flux. In each growth experiment, polycrystalline $RETa_7O_{19}$, $K_2Mo_3O_{10}$ and $B_2O_3$ were weighed in the predetermined mass ratio, hand-homogenized in an agate mortar and placed into platinum (Pt) crucibles, which were further placed into alumina crucibles covered by alumina caps (Figure 1). For the $NdTa_7O_{19}$ single-crystal growth experiments, Pt crucibles with a volume of 5 ml were used, while $ErTa_7O_{19}$ and $GdTa_7O_{19}$ growths were performed in 10 ml Pt crucibles.

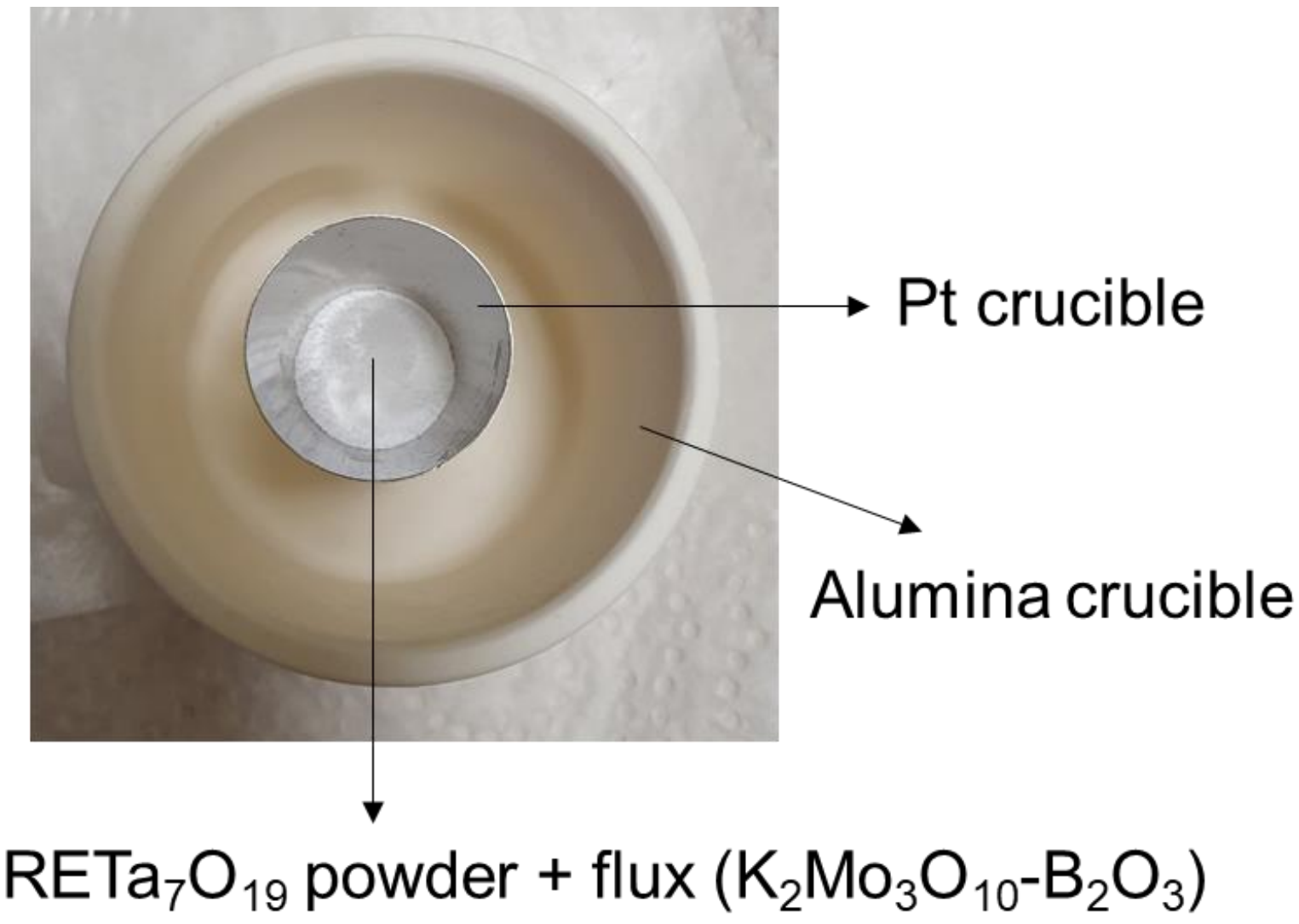

**Figure 1.** The experimental setup for single-crystal growth, showing a platinum crucible containing the reaction mixture of $NdTa_7O_{19}$ and the $K_2Mo_3O_{10}$–$B_2O_3$ flux, placed inside an alumina crucible, which was covered with an alumina cap.

In $NdTa_7O_{19}$ single-crystal growth, cooling was done in two steps; first, a slower cooling step of 0.5, 1 or 2 °C/h was employed, followed by a rapid cooling step of 10 or 50 °C/h (Table 1) with a minimal dwell time at the transition temperature in between the steps. At the end of rapid cooling, the reaction mixture was let to cool naturally. The crystals were separated from the crucible using deionized water in an ultrasonic bath.

The following growth parameters were optimized for $NdTa_7O_{19}$: **i**) batch size, **ii**) flux-to-material mass ratio, **iii**) flux composition ($K_2Mo_3O_{10}$–$B_2O_3$ mass ratio), **iv**) dwell temperature, **v**) dwell time and **vi**) cooling rate. These parameters were varied and optimized in order to grow the largest $NdTa_7O_{19}$ single crystals (Table 1). The optimal conditions were slightly modified for subsequent growth experiments of $ErTa_7O_{19}$ (Table 2). The dwell time, dwell temperature, transition temperature and rapid cooling rate were kept constant at 24 hours, 1100 °C, 950 °C and 10°C/h.

**Table 1.** A summary of tested parameters for $NdTa_7O_{19}$ flux growth.

| Parameter | Tested conditions |
|---|---|
| Batch size (g) | 0.6, 0.7, 0.9, 1, 1.1, 1.2, 1.6, 2.2, 3, 4.4 |
| Flux to material mass ratio | 5:1, 8:1, 9:1, 10:1, 15:1 |
| Flux composition ($K_2Mo_3O_{10}$–$B_2O_3$ mass ratio) | 2.3:1, 4:1, 8:1, 9:1, 99:1 |
| Dwell temperature (°C) | 700, 1100, 1150, 1200 |
| Dwell time (h) | 2, 24, 48 |
| Slower cooling rate (°C/h) | 0.5, 1, 2 |
| Transition temperature (°C) | 950, 500 |
| Rapid cooling rate (°C/h) | 10, 50 |

**Table 2.** A summary of tested parameters for $ErTa_7O_{19}$ flux growth.

| Parameter | Tested conditions |
|---|---|
| Batch size (g) | 0.8, 0.9, 1.1, 1.22, 1.35, 2.2 |
| Flux to material mass ratio | 6:1, 7:1, 8:1, 9:1, 10:1 |
| Flux composition ($K_2Mo_3O_{10}$–$B_2O_3$ mass ratio) | 3:1, 4:1, 5:1, 9:1 |
| Slow cooling rate (°C/h) | 0.3, 0.5, 0.7, 0.8. 1, 2, 5 |

Flux growth of $GdTa_7O_{19}$ was performed using a 4:1 $K_2Mo_3O_{10}$–$B_2O_3$ mass ratio, an 8:1 flux to material mass ratio and a batch size of 0.90 g. Two slow cooling rates were tested, 0.8 and 0.5 °C/h. The remaining parameters of the temperature profile were identical to those employed for $ErTa_7O_{19}$ growth.

### 2.3 Powder X-ray diffraction

Polycrystalline $RETa_7O_{19}$ samples obtained by solid-state synthesis were first examined by PXRD using a Rigaku MiniFlex600-C Benchtop X-ray Diffractometer with Cu Kα radiation in the 2θ range of 3–120° with a step size of 0.010° and a speed of 2.5°/min. The primary flux component, $K_2Mo_3O_{10}$, was also examined using the same conditions, albeit using a smaller 2θ range of 3–90°.

Phase identification was performed using SmartLab Studio II software. The Rietveld refinements were performed using the least-square method in GSAS-II.[25]

### 2.4 Single-crystal X-ray diffraction

The grown crystals were first examined under a polarizing microscope. Selected crystals of appropriate sizes were mounted on MiTeGen Dual Thickness MicroLoops with Baysilone-Paste (Bayer-Silicone, mittelviskos) and measured on a Rigaku OD XtaLAB Synergy-S Dualflex diffractometer, equipped with a PhotonJet-S microfocus Ag Kα X-ray source and an Eiger2 R CdTe 1M hybrid-photon-counting detector. *CrysAlisPro* software[26] was employed for data collection and reduction. Crystal structures were solved by *olex2.solve* and refined by *SHELXL*[27] within *OLEX2* program.[28]

### 2.5 Laue Diffraction

White beam, X-ray backscatter diffraction was used to assess the surface quality and the crystallographic orientation of the single crystals. A real-time Laue system (Laue-Camera GmbH) was used with an X-ray Seifert ID3003 generator equipped with the MWL 120 detector.

### 2.6 Scanning electron microscopy and energy dispersive X-ray spectroscopy

For the scanning electron microscopy (SEM) analyses, a few small $NdTa_7O_{19}$ single crystals were placed on a carbon tape and carbon coated using a Balzers SCD 050 sputter coater. The imaging and compositional analyses were performed on a Thermo Fisher Quanta 650 ESEM equipped with an energy-dispersive X-ray spectrometer − EDS (Oxford Instruments, AZtec Live, Ultim Max SDD 65 mm$^2$). The accelerating voltage used was 20 kV in all cases.

### 2.7 Magnetic susceptibility

The magnetic susceptibility measurements were carried out using an MPMS-XL-5 SQUID magnetometer from Quantum Design in the 2–300 K temperature range in an applied magnetic field of 1 kOe. A single crystal of $NdTa_7O_{19}$ (16 mg), $ErTa_7O_{19}$ (4 mg) and $GdTa_7O_{19}$ (2 mg) was glued to a straw with diamagnetic Apiezon N grease in different orientations to the external magnetic field. The field dependence of the isothermal magnetization was measured between −50 and 50 kOe at 2 K.

## 3. RESULTS AND DISCUSSION

### 3.1 Solid-state synthesis

The primary flux component, $K_2Mo_3O_{10}$, was synthesized using a solid-state thermal method according to Equation 1, as it is not commercially available. Rietveld refinement analysis on the resulted polycrystalline $K_2Mo_3O_{10}$ showed 82.0(4) wt% of the main phase (*C*2/c space group), 9.2(2) wt% of $K_2Mo_4O_{13}$ and 8.8(3) wt% of $K_2Mo_2O_7$ (Figure 2a). The additional molybdates were likely formed by the loss of the potassium source. This was subsequently optimized to almost pure phase, 98(1)%, $K_2Mo_3O_{10}$. However, the crystal growths were performed using the batch presented in Figure 2a.

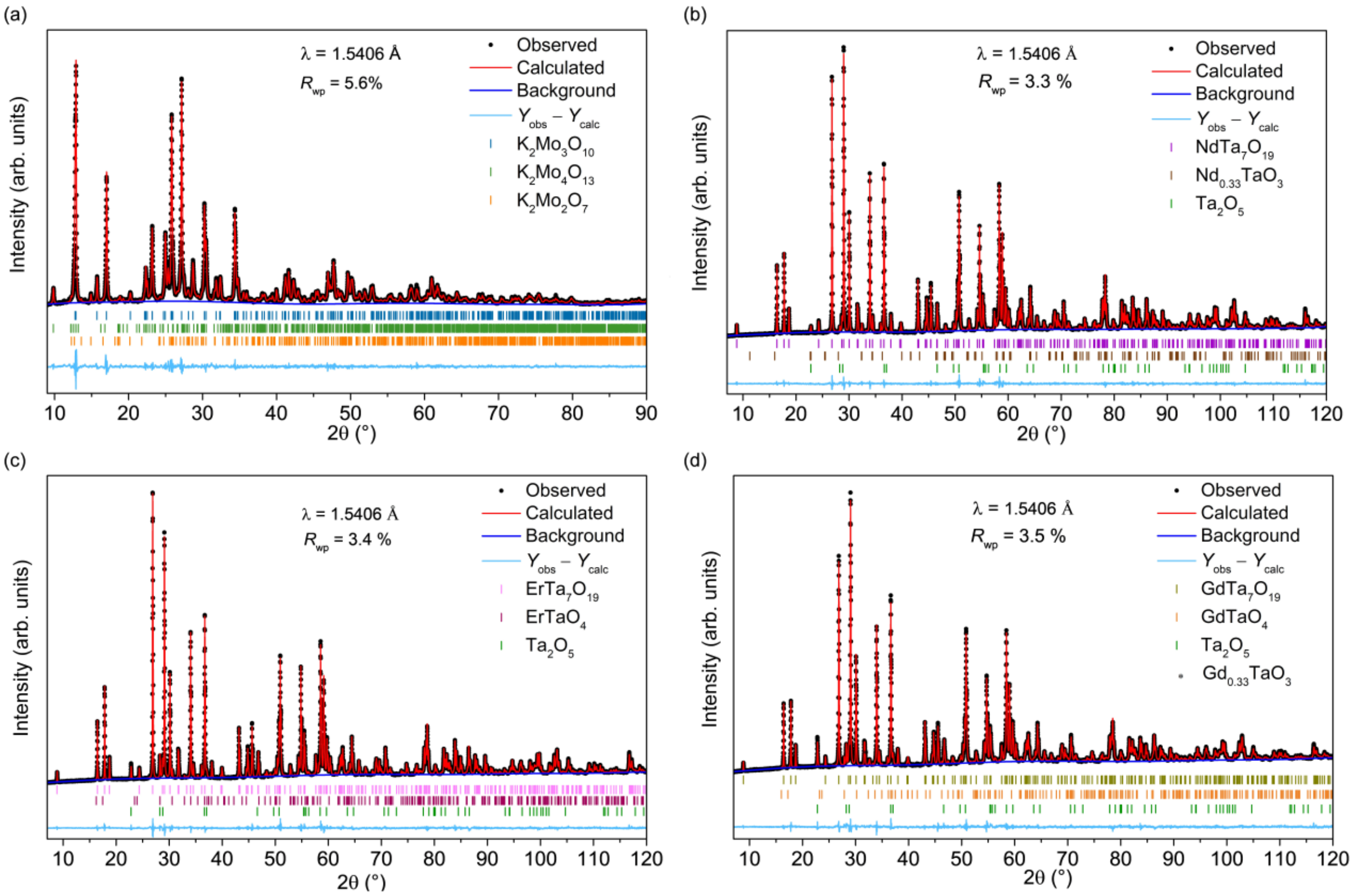

**Figure 2.** Rietveld refinement profiles of laboratory powder X-ray diffraction data (Cu Kα, $T = 293$ K) of (**a**) the primary flux component, $K_2Mo_3O_{10}$, (**b**) $NdTa_7O_{19}$, (**c**) $ErTa_7O_{19}$ and (**d**) $GdTa_7O_{19}$. The Bragg 2θ positions are marked as bars below each diffractogram.

As mentioned in the previous section, this study aimed to perform single-crystal growth experiments starting from polycrystalline $RETa_7O_{19}$ to avoid the secondary phases previously reported in the literature.[1,2,3,4] Thus, the three phases, $NdTa_7O_{19}$, $ErTa_7O_{19}$, and $GdTa_7O_{19}$ were first prepared by solid-state synthesis. The phase identifications and Rietveld refinement analyses of the powder data indicated that the three members of the series are isostructural and adopt the $P\bar{6}c2$ space group, in good agreement with the literature (Figure 2b, 2c and 2d).[29,30] Detailed results of the Rietveld refinement analysis are available in the Supplementary Information File, Table S1. A 0.6% decrease in the unit cell volume is observed from Nd ($V$ = 668.536(4) Å$^3$) to Gd ($V$ = 664.780(7) Å$^3$) and a 1% reduction to the Er member ($V$ = 661.418(5) Å$^3$), as expected following the decreasing radii of the rare-earth ions due to the lanthanoid contraction.[31] A parallel trend is observed with the magnetic intralayer RE−RE distance, which reduces from 6.2228(1) Å for Nd to 6.2119(1) Å for Gd and 6.2018(1) Å for the Er member, while the magnetic interlayer distances decrease from 9.9678(1) Å for Nd to 9.9465(1) Å for Gd and 9.9285(1) Å for the Er member.

A small fraction of secondary phases was observed in the powder X-ray diffractograms of each heptatantalate (Figure 2b, 2c and 2d), identified as unreacted $Ta_2O_5$ and two additional rare-earth tantalates. The Rietveld refinement analysis revealed 1.2(2) wt% of $Ta_2O_5$ and 1.2(1) wt% of $Nd_{0.33}TaO_3$ for the $NdTa_7O_{19}$ sample, while the $ErTa_7O_{19}$ sample contained 5.1(1) wt% of $Ta_2O_5$ and 2.2(1) wt% of $ErTaO_4$, with the $GdTa_7O_{19}$ sample containing a slighter larger content of impurities, namely 7.9(1) wt% of $Ta_2O_5$, 1.6(1) wt% of $GdTaO_4$ and 1.0(1) wt% of $Gd_{0.33}TaO_3$. One way to minimize the amount of these phases is to increase the annealing temperature in the solid-state reactions and/or the dwell time. Despite the small amount of impurities, the powders were used for single-crystal growth, as the process can also serve as a purification method, ensuring no significant impact on the study.

### 3.2 Single-crystal growth

The optimized single-crystal growth procedure employed in this study resulted in purple transparent $NdTa_7O_{19}$ crystals of predominantly hexagonal morphologies (Figure 3a), which measured up to 3.5 mm in lateral size and up to 2 mm in thickness.

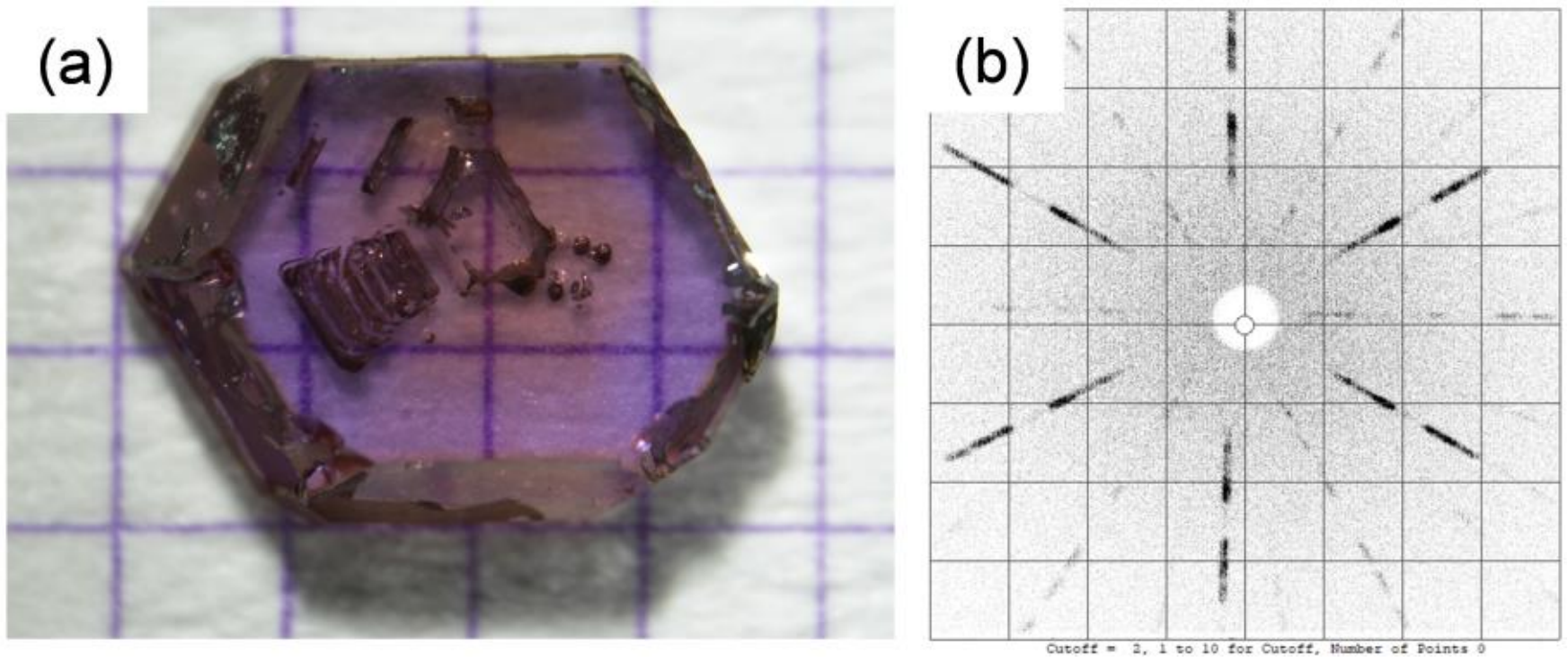

**Figure 3.** (**a**) A hexagonal $NdTa_7O_{19}$ crystal grown in this study on a millimeter grid and (**b**) the corresponding X-ray Laue back-reflection pattern observed along the *c* crystallographic direction, displaying the six-fold symmetry.

The orientation of the crystals[1,2,3,4,12]—with hexagonal planes being parallel to the *ab* plane and perpendicular to the *c* crystallographic axis (Figure 3a)—was confirmed by Laue X-ray diffraction (Figure 3b).

The optimization of the temperature profile for $NdTa_7O_{19}$ flux growth identified the dwell temperature and the slow cooling rate as pivotal parameters (Figure 4). The optimal dwell temperature was determined to be 1100 °C. While previous studies[2,4] reported higher dwell temperatures of 1150 °C and 1200 °C, in the present study, these elevated temperatures resulted in increased flux volatility, which led to the formation of numerous nucleation sites and, consequently, smaller single crystals (see Supplementary Information File, Figure S2 for pictures of the resulted crystals). The optimized cooling rate of the first step which yielded the largest crystal sizes was 0.5 °C/h. Higher cooling rates of 1 or 2 °C/h resulted in multiple smaller crystals. The dwell time was found to be a less significant parameter; observations indicated that a 48-hour dwell time did not result in substantially different crystal sizes compared to a shorter 24-hour dwell time.

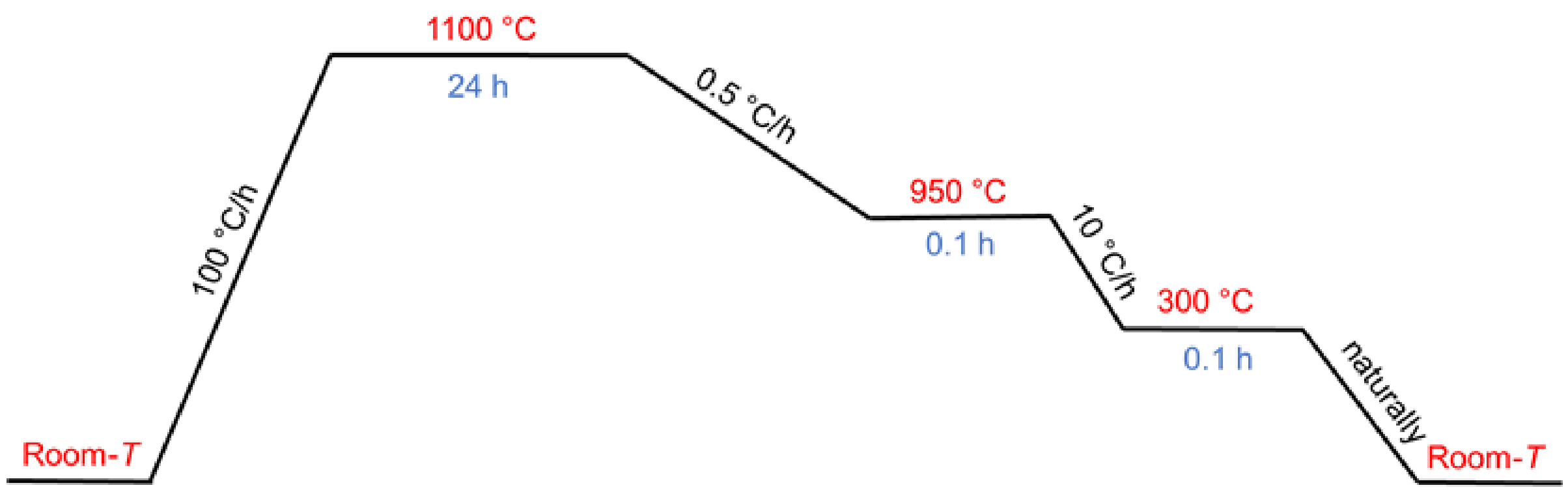

**Figure 4.** The optimal temperature profile for flux growth of $NdTa_7O_{19}$ using the $K_2Mo_3O_{10}$–$B_2O_3$ flux. The dwell temperature, 1100 °C, and the cooling rate of the first step, 0.5 °C/h, were identified as critical crystal-growth parameters.

Other parameters tested during the growth process (displayed in Table 1) included: (**i**) the batch size, (**ii**) the flux-to-material mass ratio and (**iii**) the flux composition.

(**i**) The optimal batch size was determined to be approximately 1 g. Attempts to increase crystal sizes by doubling or further increasing the mass proved ineffective, resulting only in the formation of multiple smaller crystals.

(**ii**) The optimal range of flux-to-material mass ratios was found to be between 8:1 and 10:1. Other ratios, including 5:1 and 15:1, were also investigated, however, they did not yield any large single crystals.

(**iii**) The flux composition also turned out to be a crucial parameter, being the primary factor affecting the crystal morphology, while additionally influencing the sizes. Experiments with $K_2Mo_3O_{10}$–$B_2O_3$ mass ratio of 9:1 resulted in plate-like crystals. An increase of the $B_2O_3$ content in the $K_2Mo_3O_{10}$–$B_2O_3$ mass ratio led to increased thicknesses of the crystals, transitioning from plate-like towards isometric morphologies (Supplementary Information, Figure S1), which was also observed in previous literature reports.[2,3,4]

The optimal $K_2Mo_3O_{10}$–$B_2O_3$ mass ratio of 4:1, yielded predominantly hexagonal crystals of approximately 1 mm thickness. Elongated morphologies, as previously reported in the literature,[2,3,4] were not observed. Further experiments with higher or lower ratios (Table 1), failed to produce large single crystals (Figure S2). The optimal parameters for $NdTa_7O_{19}$ growth are summarized below (Table 3).

**Table 3.** The optimal growth parameters for $NdTa_7O_{19}$ growth.

| Parameter | Optimal conditions |
|---|---|
| Batch size (g) | 0.9−1.1 |
| Flux to material mass ratio | 8:1−10:1 |
| Flux composition ($K_2Mo_3O_{10}$–$B_2O_3$ mass ratio) | 4:1 |
| Dwell temperature | 1100 °C |
| Slow cooling rate | 0.5 °C/h |

Seeding, a key parameter in crystal growth, was also investigated in this study. A small crystal ("seed") obtained from a previous experiment was placed at the bottom of the crucible. The role of the seed crystal is to hinder multiple nucleation from occurring.[32] This method promotes controlled nucleation and the growth of a single crystal with fewer defects and better quality. However, seeding had minimal impact on the crystal size, indicating it is not a critical factor in this growth process.

The optimal $NdTa_7O_{19}$ crystal growth conditions were also employed for the growth of $ErTa_7O_{19}$ single crystals, with some optimization of the cooling rates (Table 2). The largest crystals were grown using a cooling rate of 0.8 °C/h, while lower rates of 0.5 and 0.3 °C/h resulted in increased flux volatility, as deduced from the mass loss.

The flux-to-material mass ratios that yielded the largest crystal sizes were 7:1 and 8:1, while the optimal batch size was found to be 0.8−0.9 g. Higher masses resulted in the formation of multiple smaller crystals. The optimal $K_2Mo_3O_{10}$–$B_2O_3$ mass ratios ranged from 3:1 to 5:1. Similarly as in the $NdTa_7O_{19}$ growth, the crystal morphology was influenced by the flux composition. Specifically, a higher content of $B_2O_3$ in the $K_2Mo_3O_{10}$–$B_2O_3$ flux resulted in decreased lateral sizes of the crystals and increased thicknesses. The optimal growth parameters are summarized in Table 4.

Pink $ErTa_7O_{19}$ crystals exhibited predominantly hexagonal morphologies (Figure 5a) with lateral sizes up to 2.5 mm and thicknesses up to 0.5 mm. The orientation of $ErTa_7O_{19}$ crystals (Figure 5b) was confirmed by X-ray Laue diffraction (Figure 5c).

**Table 4.** The optimal parameters identified for the $ErTa_7O_{19}$ growth.

| Parameter | Optimal conditions |
|---|---|
| Batch size (g) | 0.8−0.9 |
| Flux to material mass ratio | 7:1−8:1 |
| Flux composition ($K_2Mo_3O_{10}$–$B_2O_3$ mass ratio) | 3:1−5:1 |
| Dwell temperature | 1100 °C |
| Slow cooling rate | 0.8 °C/h |

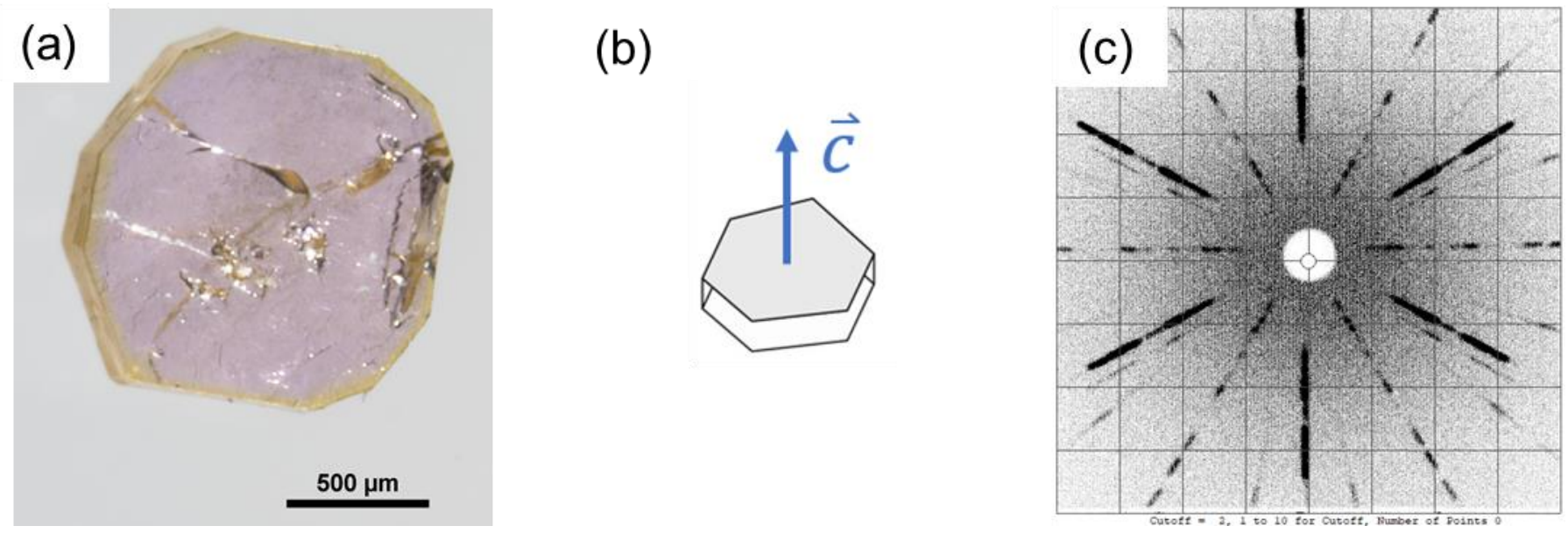

**Figure 5.** (**a**) An $ErTa_7O_{19}$ single crystal grown by the flux method. (**b**) The expected orientation of the crystallographic *c*-axis of an $ErTa_7O_{19}$ crystal. (**c**) Corresponding X-ray Laue back-reflection pattern collected on an $ErTa_7O_{19}$ single crystal showing the six-fold axis along the *c* crystallographic direction.

For $GdTa_7O_{19}$ single-crystal growth, the optimal $NdTa_7O_{19}$ growth conditions were employed as listed in Table 3. Similar to the other two representatives of the series, yellow $GdTa_7O_{19}$ single crystals exhibited pseudo-hexagonal morphologies with lateral sizes up to 4 mm and thicknesses up to 0.5 mm (Figure 6 a−c).

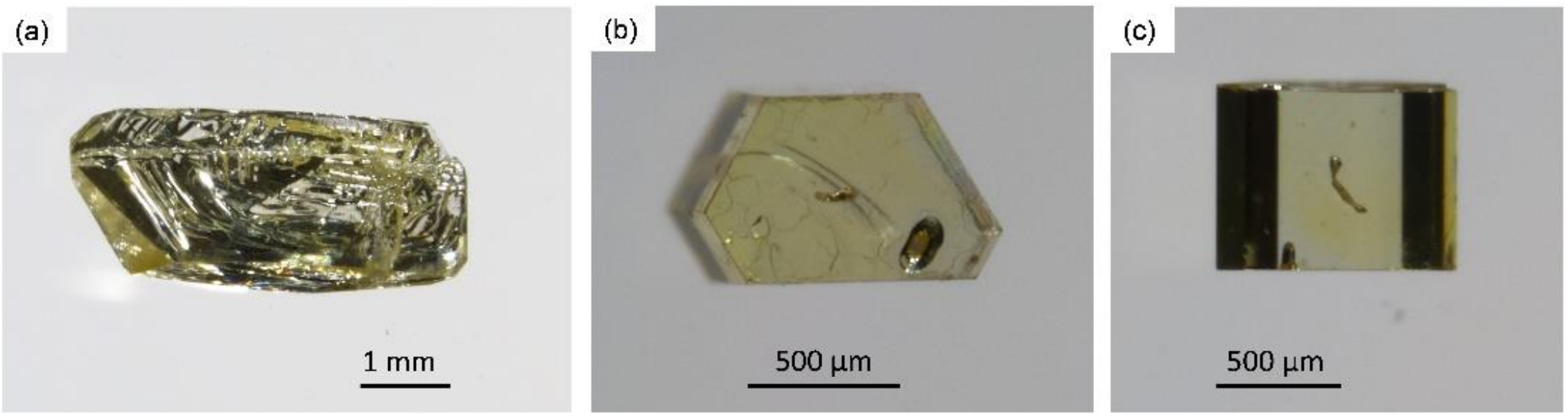

**Figure 6.** (**a**) The largest as-grown pseudo-hexagonal $GdTa_7O_{19}$ crystal. (**b**) Frontal view and (**c**) side view of another, smaller hexagonal $GdTa_7O_{19}$ single crystal grown in this study.

### 3.3 Chemical analysis

The grown single crystals of $NdTa_7O_{19}$, $ErTa_7O_{19}$ and $GdTa_7O_{19}$ were analyzed by PXRD and single-crystal XRD (SCXRD). All structural information can be found in the Supplementary Information File, Tables S1–S6. All the structures were solved and refined in the $P\bar{6}c2$ space group.[29] Initially, the $P6_3/mcm$ space group was identified for $CeTa_7O_{19}$ and the $RETa_7O_{19}$ series.[33] However, a year later, the same author reported a preference for the $P\bar{6}c2$ space group.[34]

The RE ions form a perfect triangular lattice (side view: Figure 7a, top view: Figure 7b), where each RE ion is coordinated by eight oxygen ions, forming distorted $REO_8$ polyhedra that share edges with neighboring $TaO_6$ octahedra. The magnetic layers are separated by two nonmagnetic Ta-layers composed of $TaO_7$ polyhedra.

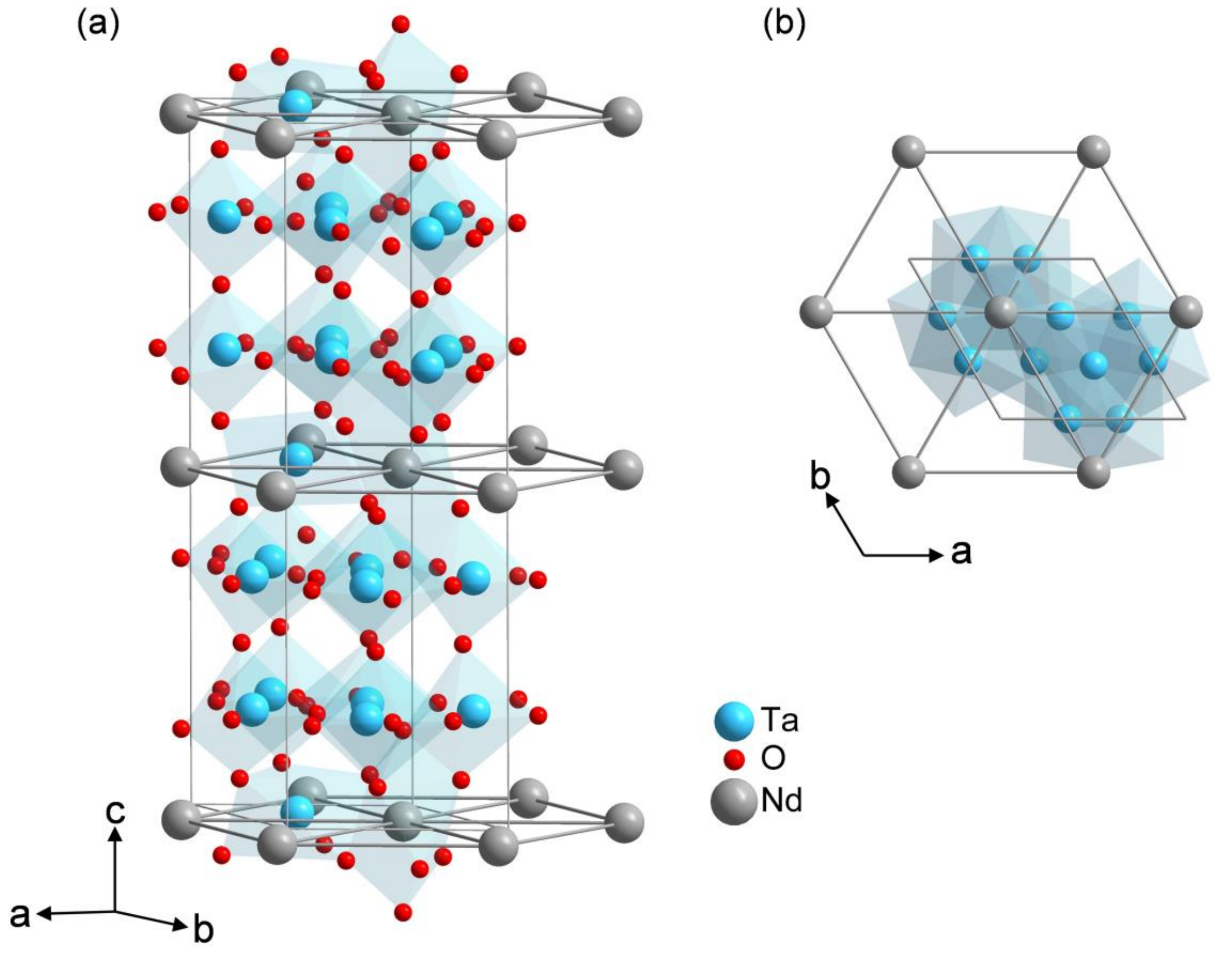

**Figure 7.** Crystal structure of $RETa_7O_{19}$ (here RE = Nd) (**a**) in the *ab*-plane and (**b**) along the *c* crystallographic axis, showing the perfect triangular lattice within the magnetic layers, separated by two nonmagnetic layers of $TaO_7$ polyhedra. The oxygen atoms in (**b**) are omitted for clarity.

A systematic decrease in the unit cell volume (Supplementary Information File, Table S2), the magnetic interlayer distances and the RE−RE distances is observed from RE = Nd to Er, following the decrease in the ionic radii (Supplementary Information File, Tables S4–6).[31] These findings are consistent with the observations on the polycrystalline heptatantalates (see Section 3.1 / Supplementary Information File, Table S1).

The SEM-EDS point analysis was employed for elemental analysis and to identify possible flux inclusions. A representative SEM-EDS point analysis spectrum of a $NdTa_7O_{19}$ crystal, along with the corresponding SEM image of the crystal (Figure 8a), shows peaks of all the constituent elements without any flux components or other impurities. The average atomic percent values obtained from 11 point analyses are: 3.7(1)% Nd, 26.0(1)% Ta and 70.4(1)% O, perfectly corresponding to the theoretical ratio 1/27:7/27:19/27 and thereby confirming the stoichiometry of the sample. A few dark spots observed on the surface correspond to particles of the main flux component, $K_2Mo_3O_{10}$.

The EDS point analyses of $ErTa_7O_{19}$ (Figure 8b) and $GdTa_7O_{19}$ single crystals (Figure 8c) confirm the presence of Er, Gd, Ta and O. For $ErTa_7O_{19}$, the average elemental composition from 10 point analyses is 3.5(2) at% Er, 25.8(1) at% Ta, 70.4 at(1)% O with trace amounts of K, 0.1(1) at%, and Mo, 0.2(1) at%, likely from the flux on the surface. The $GdTa_7O_{19}$ crystal shows slightly higher flux residues (Figure 8c) with 7 point analyses yielding 3.5(3) at% Gd, 21.6(7) at% Ta, 67.1(3) at% O, 6.3(4) at% K and 1.5(6) at% Mo.

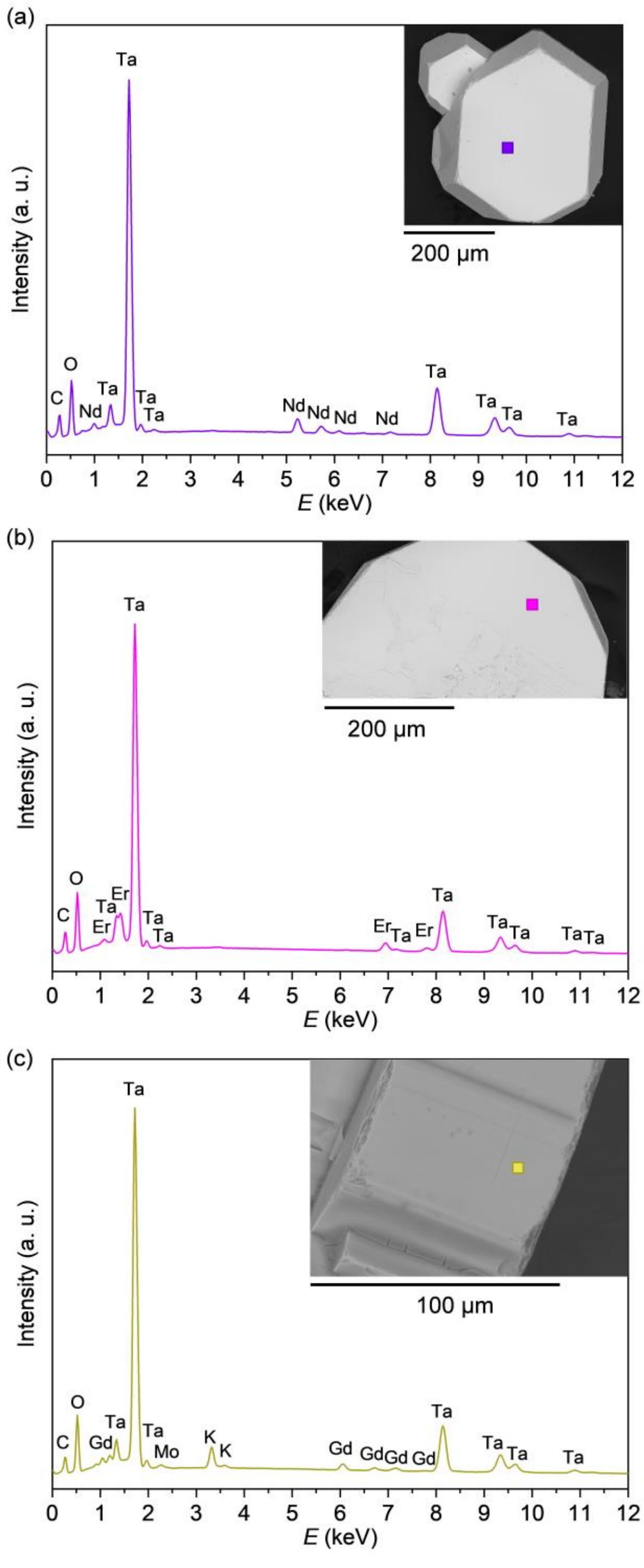

**Figure 8.** (**a**) SEM-EDS spectrum of a hexagonal $NdTa_7O_{19}$ crystal, with the analysis area marked by a violet square, confirming stoichiometric Nd, Ta, and O ratios. (**b**) SEM-EDS spectrum of an $ErTa_7O_{19}$ crystal, with the analysis area indicated by a pink square, confirming Er, Ta, and O. (**c**) SEM-EDS spectrum of a $GdTa_7O_{19}$ crystal, with the analysis area marked by a yellow square (inset), detecting Gd, Ta, O, and traces of K and Mo (from the residual flux on the surface).

### 3.4 Magnetic properties

The magnetic susceptibility of $NdTa_7O_{19}$ (Figure 9a) shows no signs of magnetic ordering down to 2 K, which suggests a possible dynamical magnetic ground state and is consistent with previous results obtained on polycrystalline $NdTa_7O_{19}$.[5]

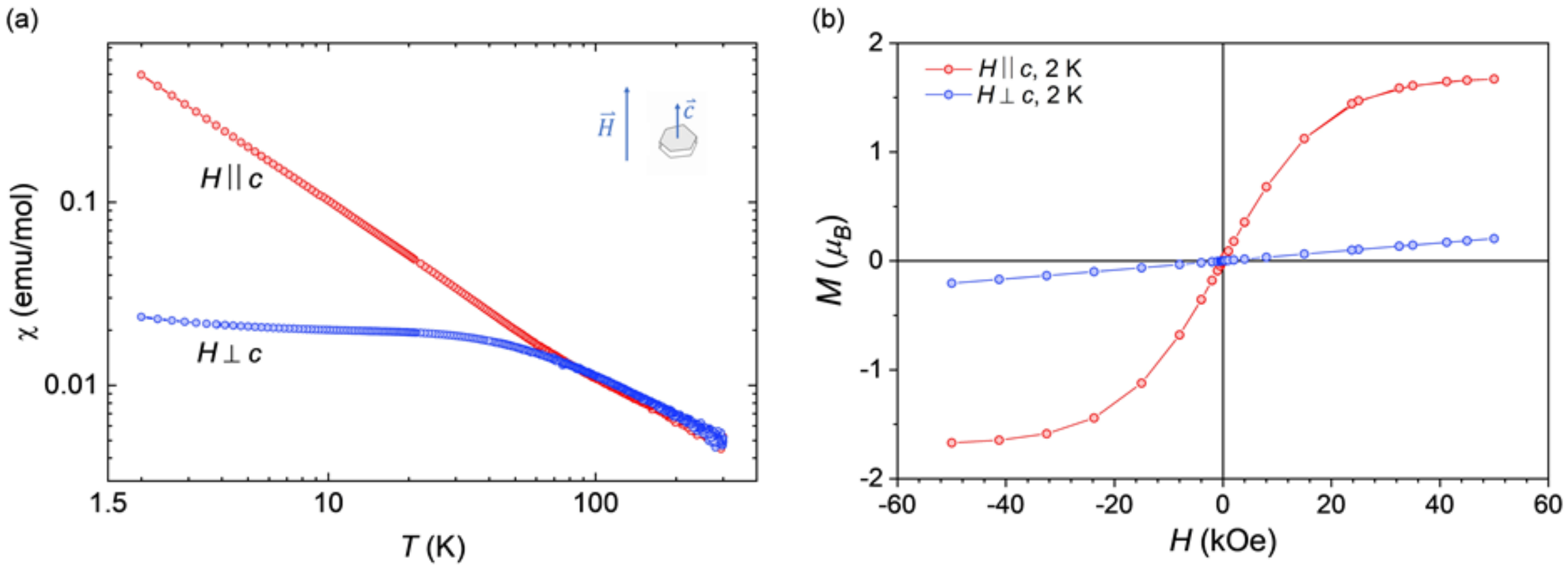

**Figure 9.** Magnetic susceptibility of $NdTa_7O_{19}$ single crystal (**a**) and *M*(*H*) curves (**b**), showing a pronounced difference in the magnetic response in both principal crystallographic directions, $H \parallel c$ and $H \perp c$, which represents direct evidence of the easy-axis magnetic anisotropy.

When the magnetic field *H* of 1 kOe is applied parallel ($H \parallel c$) and perpendicular ($H \perp c$) to the *c*-axis (Figure 9b), a very similar paramagnetic response is observed for both crystallographic directions as the temperature decreases. However, below 90 K, the magnetic responses of the two orientations become increasingly different in size and behavior. This divergence is evident in both the magnetic susceptibility (Figure 9a) and field-dependent magnetization at 2 K, *M*(*H*) (Figure 9b), providing clear evidence of easy-axis single-ion magnetic anisotropy, as previously suggested by crystal-electric-field modeling of polycrystalline samples.[5]

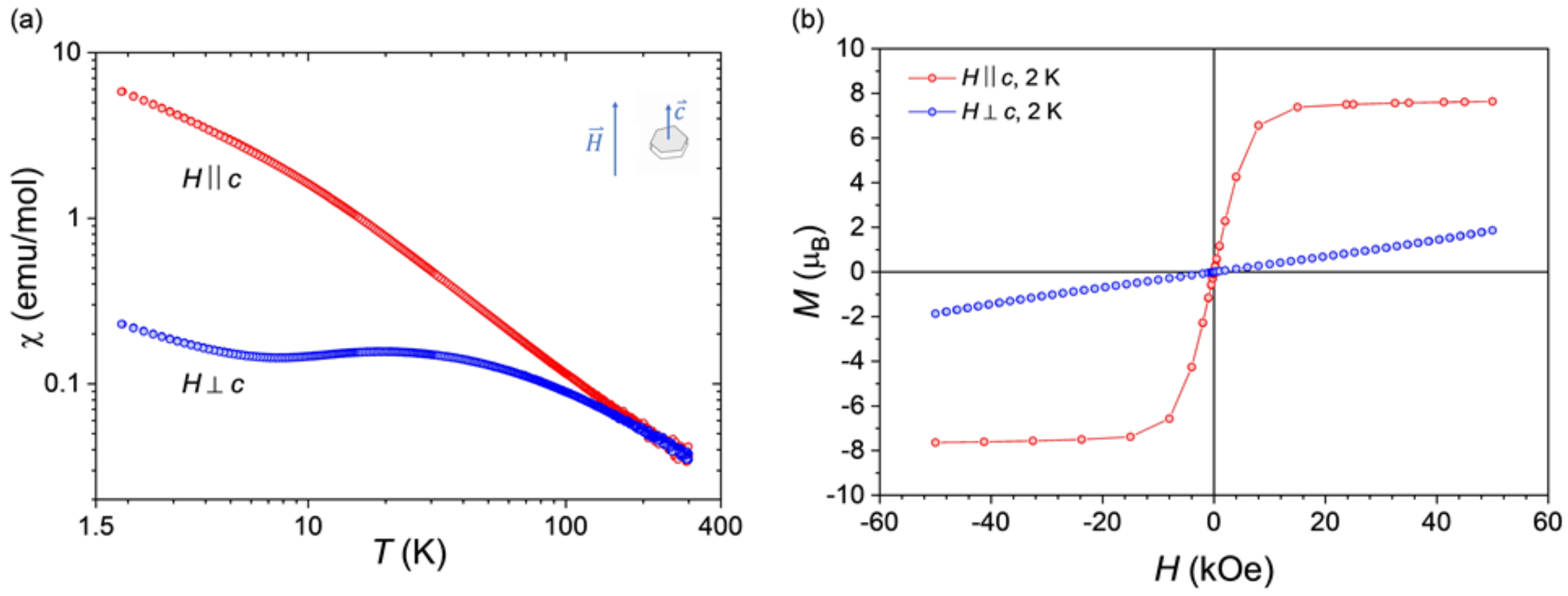

**Figure 10.** (**a**) Magnetic susceptibility of $ErTa_7O_{19}$ single crystal, showing a significant difference in the magnetic response in the crystallographic directions $H \parallel c$ and $H \perp c$ as a consequence of large magnetic anisotropy. (**b**) $M(H)$ curves for $H \perp c$ and $H \parallel c$ at 2 K. A significant magnetic anisotropy is also observed, similar to (**a**).

A similar magnetic response to $NdTa_7O_{19}$ is also observed in magnetic susceptibility and field-dependent magnetization of $ErTa_7O_{19}$ (Figure 10a and 10b). In the external field of 1 kOe, a paramagnetic behavior is observed in both crystallographic directions at high temperatures to about 150 K. However, as the temperature decreases to 2 K, a significant divergence in the magnetic response for both crystallographic directions is observed, $H \parallel c$ and $H \perp c$, again providing evidence of easy-axis type magnetic anisotropy. The absence of magnetic ordering down to 2 K suggests a similar dynamic magnetic ground state as previously observed for $NdTa_7O_{19}$.[5]

The magnetic response of $GdTa_7O_{19}$ to a field of 1 kOe (Figure 11a) is substantially different from the previous two representatives of the $RETa_7O_{19}$ series and corresponds to that of an isotropic paramagnet. The results of susceptibility as well as field-dependent magnetization (Figure 11b) are very similar for both crystallographic directions, showing the absence of anisotropy in the magnetism of this compound. This isotropic magnetic behavior aligns with a recent study,[35] which reported rare-earth dependent magnetic properties for six polycrystalline $RETa_7O_{19}$ compounds RE = Pr, Sm, Eu, Gd, Dy, Ho.

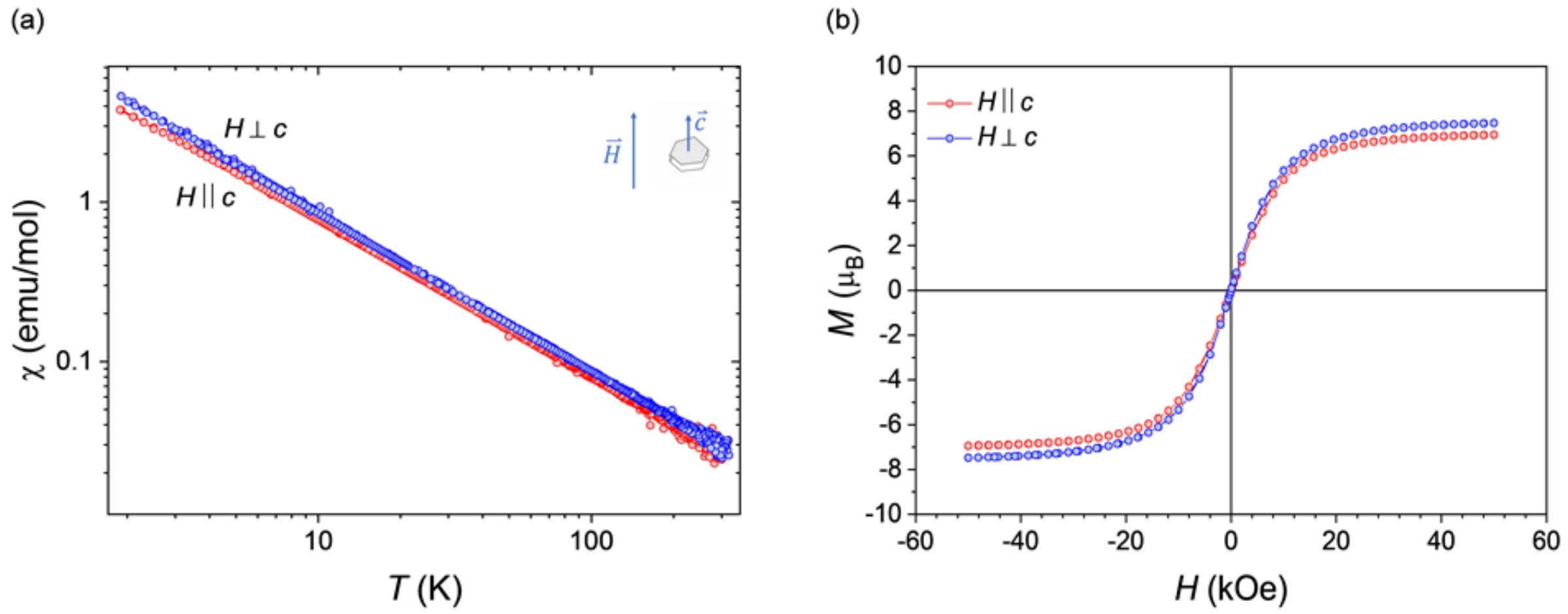

**Figure 11.** (**a**) The magnetic susceptibility of $GdTa_7O_{19}$ single crystal showing the absence of anisotropy, *i.e.*, a nearly identical paramagnetic response for both $H \parallel c$, and $H \perp c$, typical of a paramagnet. (**b**) The $M(H)$ curves for $H \parallel c$ and $H \perp c$ at 2 K, showing a similar magnetic response in both directions, in line with observations in (**a**).

## 4. CONCLUSIONS

High-quality single-crystals of $NdTa_7O_{19}$, the first Ising quantum spin liquid candidate on a triangular lattice, were successfully grown using the $K_2Mo_3O_{10}$–$B_2O_3$ flux. With lateral sizes up to 3.5 mm, thicknesses up to 2 mm, and hexagonal, plate-like morphologies, these crystals are the largest reported to date.

To maximize the yield and reduce the formation of secondary neodymium phases, the crystals were grown from polycrystalline $NdTa_7O_{19}$, obtained by optimizing a previously reported solid-state method.[24] This optimized flux method was further extended to grow large, high-quality single crystals of two additional compounds in the series, $ErTa_7O_{19}$ and $GdTa_7O_{19}$.

A substantial easy-axis magnetic anisotropy in $NdTa_7O_{19}$ was observed in its magnetic susceptibility at low temperatures down to 2 K, providing clear proof of the anisotropy indicated by earlier modelling of the crystal electric field in the polycrystalline sample.[5] Similarly, $ErTa_7O_{19}$ also exhibited a pronounced magnetic anisotropy of the same type. No evidence of magnetic ordering is found down to 2 K, suggesting a possible exotic magnetic ground state also for this representative. In contrast, $GdTa_7O_{19}$ showed a paramagnetic behavior, in agreement with findings from polycrystalline samples.[35]

The obtained high-quality single crystals will allow new insight into the exciting magnetism of these frustrated magnets. More detailed and $q$-resolved investigations of the magnetic ground state, magnetic correlations and excitations are now possible. Furthermore, the flux method holds the potential for successful growth of additional rare-earth heptatantalates in the series.

## ASSOCIATED CONTENT

**Supporting Information**. Summary of single-crystal X-ray diffraction data, data collection and structure refinement details; structural parameters derived from Rietveld refinement fits of polycrystalline $NdTa_7O_{19}$, $ErTa_7O_{19}$ and $GdTa_7O_{19}$ laboratory PXRD data; additional images of $NdTa_7O_{19}$ single crystals (PDF).

## AUTHOR INFORMATION

**Corresponding Author**

* mirela.dragomir@ijs.si

**Funding Sources**

This work was supported by the Slovenian Research and Innovation Agency (Young Researcher's program, Program P1-0125, P2-0348 and Project No. J1-50008), the Marie Skłodowska-Curie Individual Fellowship (Grant No. 101031415) and the European Research Council Starting Grant (Grant No. 950625) under the European Union's Horizon 2020 Research and Innovation Program.

## ACKNOWLEDGMENT

Assoc. Prof. Mario Novak and Dr. Mirta Herak from University of Zagreb are acknowledged for Laue diffraction measurements and valuable discussions.

Supporting Information File

# Optimized flux single-crystal growth of the quantum spin liquid candidate $NdTa_7O_{19}$ and other rare-earth heptatantalates, $ErTa_7O_{19}$ and $GdTa_7O_{19}$

Lia Šibav[1,2], Matic Lozinšek[1,2], Zvonko Jagličić[3,4], Tina Arh[1,5], Panchanana Khuntia[6,7], Andrej Zorko[1,5] and Mirela Dragomir[1,2*]

*[1]Jožef Stefan Institute, Jamova cesta 39, 1000, Ljubljana, Slovenia*

*[2]Jožef Stefan International Postgraduate School, Jamova cesta 39, 1000, Ljubljana, Slovenia*

*[3]Faculty of Civil and Geodetic Engineering, University of Ljubljana, Jamova cesta 2, 1000 Ljubljana, Slovenia*

*[4]Institute of Mathematics, Physics and Mechanics, Jadranska cesta 19, 1000 Ljubljana, Slovenia*

*[5]Faculty of Mathematics and Physics, University of Ljubljana, 1000, Ljubljana, Slovenia*

*[6]Department of Physics, Indian Institute of Technology Madras, Chennai 600036, India*

*[7]Quantum Centre of Excellence for Diamond and Emergent Materials, Indian Institute of Technology Madras, Chennai 600036, India*

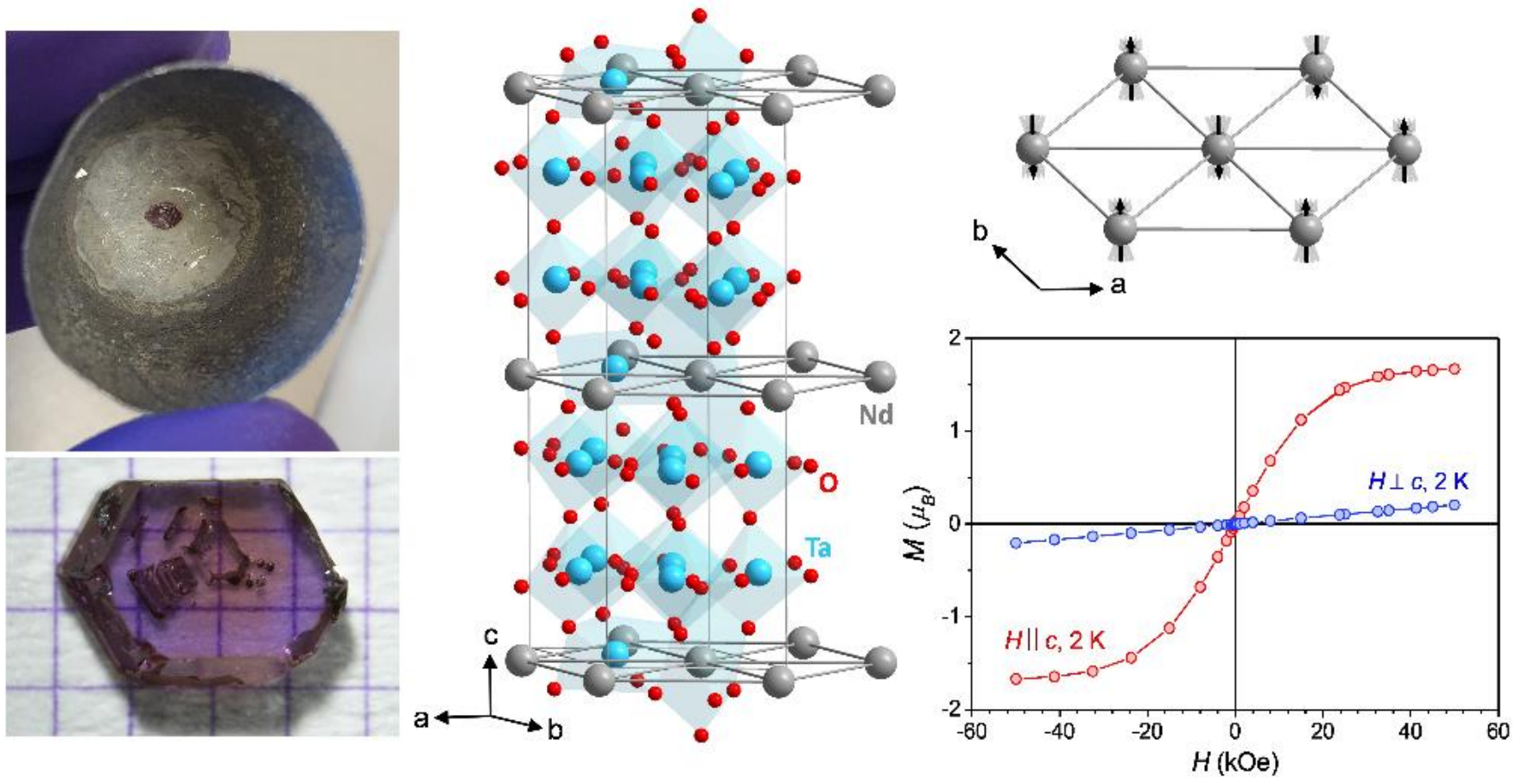

**Table S1.** A summary of Rietveld refinement analysis results on the polycrystalline $NdTa_7O_{19}$, $ErTa_7O_{19}$ and $GdTa_7O_{19}$. The data was collected at room-temperature using CuKα radiation (λ = 1.5406 Å).

| | **$NdTa_7O_{19}$** | **$GdTa_7O_{19}$** | **$ErTa_7O_{19}$** |
|---|---|---|---|
| **Space group** | $P\bar{6}c2$ | $P\bar{6}c2$ | $P\bar{6}c2$ |
| ***a, b* (Å)** | 6.22277(3) | 6.21188(4) | 6.20178(3) |
| ***c* (Å)** | 19.93551(5) | 19.89306(8) | 19.85693(5) |
| ***V* (Å$^3$)** | 668.536(4) | 664.780(6) | 661.418(5) |
| **$R_{wp}$ (%)** | 3.30 | 3.45 | 3.43 |
| **$R_{exp}$ (%)** | 2.32 | 2.35 | 2.30 |
| **$S$ ($R_{wp}/R_{exp}$)** | 1.42 | 1.47 | 1.49 |
| **Atom** | *x*/*y*/*z*<br>*U*iso | *x*/*y*/*z*<br>*U*iso | *x*/*y*/*z*<br>*U*iso |
| ***R*** | 0.66667/0.33333/0<br>0.0036(6) | 0.66667/0.33333/0<br>0.0188(16) | 0.66667/0.33333/0<br>0.0057(7) |
| **Ta1** | 0.33333/0.66667/0<br>0.0050 | 0.33333/0.66667/0<br>0.0023(8) | 0.33333/0.66667/0<br>0.0050 |
| **Ta2** | 0.63963(25)/0.63957(30)/0.15589(1)<br>0.00318(19) | 0.64033(26)/0.63831(29)/0.15576(2)<br>0.00666(26) | 0.64019(28)/0.63793(31)/0.15550(1)<br>0.00353(22) |
| **O1** | 0.7391(28)/0.977(5)/0.15315(19)<br>0.0133(15) | 0.7372(27)/0.973(5)/0.15324(26)<br>0.0171(21) | 0.7305(28)/0.958(5)/0.15198(20)<br>0.0190(18) |
| **O2** | 0.5688(8)/0.6210(10)/0.05810(18)<br>0.0050 | 0.5690(14)/0.6051(15)/0.05869(23)<br>0.0127(23) | 0.5747(12)/0.6121(14)/0.05700(19)<br>0.0104(18) |
| **O3** | 0.6216(14)/0.5838(14) /0.2500<br>0.0050 | 0.6240(23)/0.5889(25)/0.2500<br>0.015(4) | 0.6187(16)/0.5803(15)/0.2500<br>0.0050 |
| **O4** | 0.33333/0.66667/0.1639(5)<br>0.0050 | 0.33333/0.66667/0.1658(5)<br>0.0050 | 0.33333/0.66667/0.1639(5)<br>0.0050 |
| **O5** | 0.66667/0.33333/0.1301(6)<br>0.0068(31) | 0.66667/0.33333/0.1263(8)<br>0.022(5) | 0.66667/0.33333//0.1271(6)<br>0.006(3) |

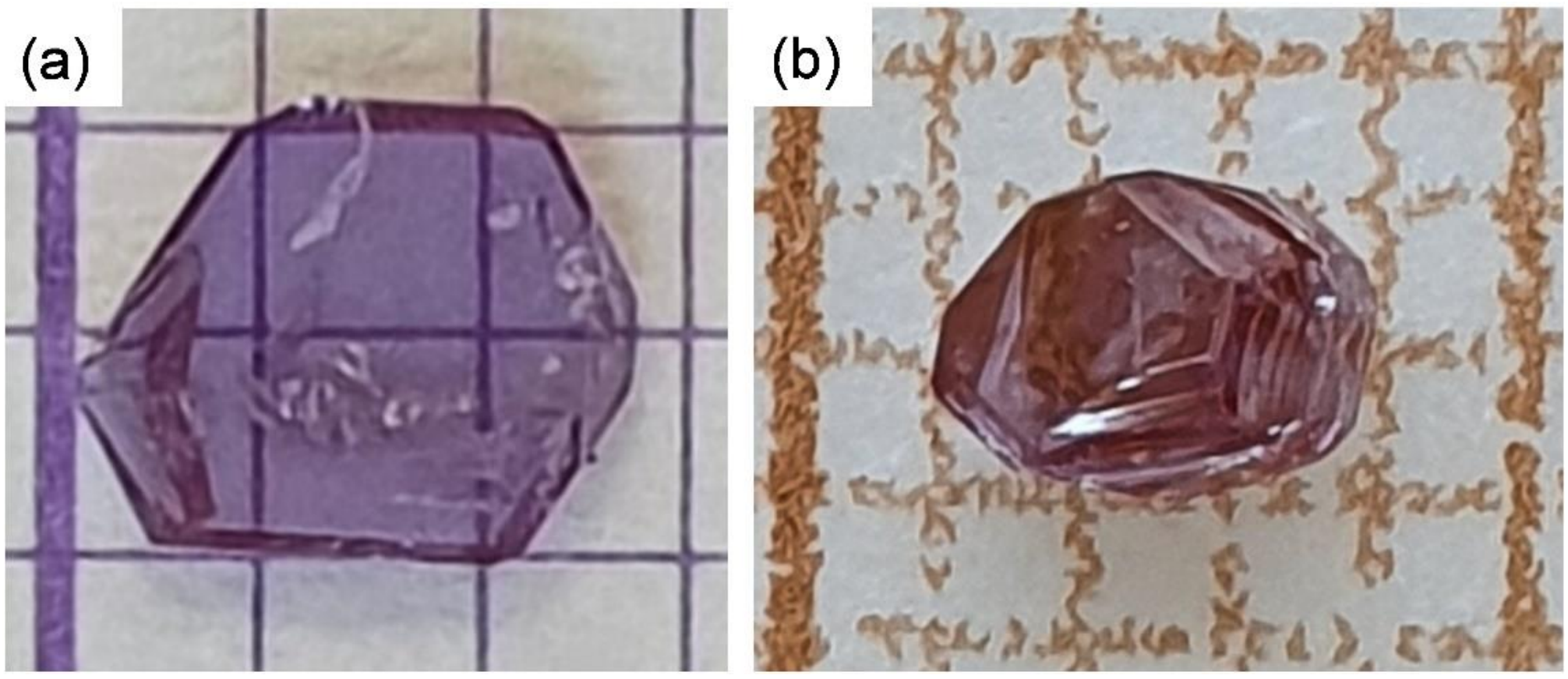

**Figure S1.** Single crystals of $NdTa_7O_{19}$. (**a**) A plate-like crystal grown with a 9:1 $K_2Mo_3O_{10}/B_2O_3$ flux mass ratio. (**b**) A crystal with larger thickness, grown with a 4:1 $K_2Mo_3O_{10}/B_2O_3$ mass ratio.

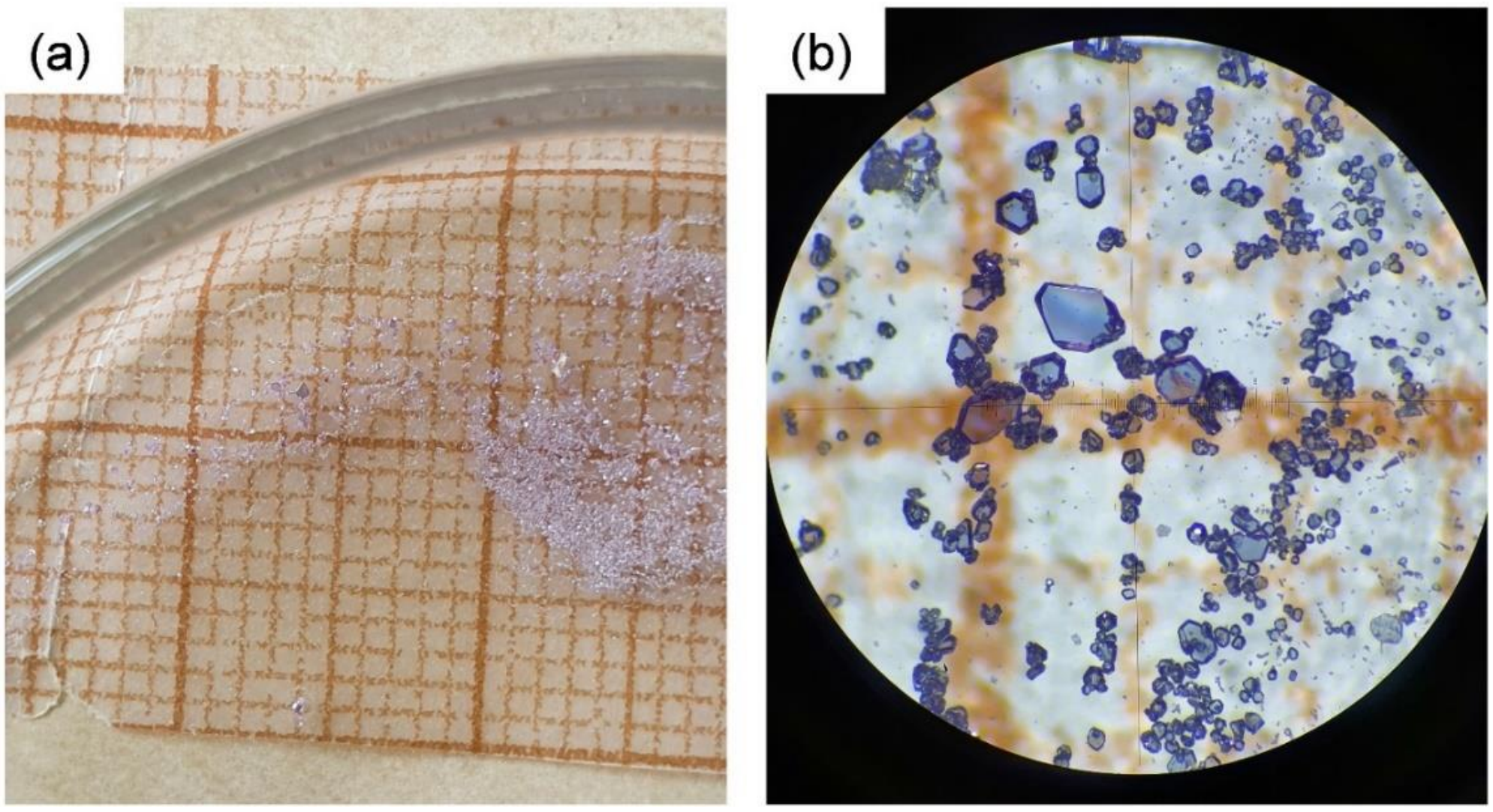

**Figure S2.** Representative images of small $NdTa_7O_{19}$ single crystals, grown in the process of optimization that did not yield crystals of significant size: (**a**) displayed on a millimeter grid for size reference and (**b**) in a magnified view on the millimeter grid by an optical microscope, highlighting the crystal morphology.

**Table S2.** Summary of single-crystal X-ray diffraction crystal data, data collection and structure refinement details for the crystal structures of $NdTa_7O_{19}$, $GdTa_7O_{19}$, and $ErTa_7O_{19}$. Collected at 100K

| **Compound** | **$NdTa_7O_{19}$** | **$GdTa_7O_{19}$** | **$ErTa_7O_{19}$** |
|---|---|---|---|
| $M_r$ | 1714.89 | 1727.90 | 1737.91 |
| Crystal system | hexagonal | hexagonal | hexagonal |
| Space group | $P\bar{6}c2$ | $P\bar{6}c2$ | $P\bar{6}c2$ |
| $a$ [Å] | 6.21494(10) | 6.20871(10) | 6.19718(6) |
| $b$ [Å] | 6.21494(10) | 6.20871(10) | 6.19718(6) |
| $c$ [Å] | 19.9382(4) | 19.9087(4) | 19.8613(2) |
| $\alpha$ [°] | 90 | 90 | 90 |
| $\beta$ [°] | 90 | 90 | 90 |
| $\gamma$ [°] | 120 | 120 | 120 |
| $V$ [Å$^3$] | 666.94(3) | 664.62(3) | 660.581(16) |
| $Z$ | 2 | 2 | 2 |
| $\rho_{calc}$ [g/cm$^3$] | 8.539 | 8.634 | 8.737 |
| Crystal size [mm$^3$] | 0.090 × 0.077 × 0.050 | 0.208 × 0.135 × 0.029 | 0.160 × 0.139 × 0.077 |
| Radiation type | Ag Kα | Ag Kα | Ag Kα |
| $\lambda$ [Å] | 0.56087 | 0.56087 | 0.56087 |
| $T$ [K] | 100 | 100 | 100 |
| $\mu$ [mm$^{-1}$] | 32.772 | 33.482 | 34.408 |
| $F$(000) | 1446 | 1454 | 1462 |
| $\theta_{max}$ [°] | 32.324 | 32.245 | 30.653 |
| Index ranges | $-11 \leq h \leq 11$<br>$-9 \leq k \leq 11$<br>$-36 \leq l \leq 35$ | $-11 \leq h \leq 10$<br>$-11 \leq k \leq 9$<br>$-35 \leq l \leq 37$ | $-11 \leq h \leq 10$<br>$-11 \leq k \leq 11$<br>$-36 \leq l \leq 35$ |
| Reflections collected | 13542 | 16495 | 40463 |
| Independent reflections | 1471 | 1525 | 1431 |
| Reflections with [$I > 2\sigma(I)$] | 1432 | 1462 | 1374 |
| $R_{int}$ | 0.0473 | 0.0573 | 0.1374 |
| $R_{sigma}$ | 0.0231 | 0.0249 | 0.0335 |
| Data/restraints/parameters | 1471/0/45 | 1525/6/45 | 1431/6/46 |
| $S$ | 1.174 | 1.158 | 1.201 |
| $R_1$, $wR_2$ [$I > 2\sigma(I)$] | 0.0198, 0.0447 | 0.0196, 0.0408 | 0.0237, 0.0524 |
| $R_1$, $wR_2$ [all data] | 0.0208, 0.0450 | 0.0213, 0.0415 | 0.0254, 0.0545 |
| $\Delta\rho_{min}$, $\Delta\rho_{max}$ [eÅ$^{-3}$] | −1.528, 2.546 | −2.330, 2.121 | −3.816, 3.301 |
| Flack $x$ | −0.010(15) | 0.007(14) | BASF = 0.63(4) [a] |

[a] refined as an inversion twin

CSD Deposition Number 2407250 (for $NaTa_7O_{19}$), 2407251 (for $GdTa_7O_{19}$), and 2407252 (for $ErTa_7O_{19}$) contain the supplementary crystallographic data for this paper. These data can be obtained free of charge from FIZ Karlsruhe via www.ccdc.cam.ac.uk/structures.

The anisotropic thermal ellipsoid of atom O1 in crystal structures of $GdTa_7O_{19}$, and $ErTa_7O_{19}$ was restrained with ISOR 0.001 to prevent it from becoming non-positive definite.

**Note**: the checkCIF report two alerts level A for the $NaTa_7O_{19}$ and four for $ErTa_7O_{19}$. These are caused by large residual electron density minima and maxima. Full single-crystal datasets have been measured on several different crystals of each compounds. The aforementioned issues were observed consistently in all analysed single crystals. Typically, the problems were exacerbated in larger crystals, which indicates possible absorption effects.

**Table S3.** Fractional atomic coordinates and isotropic or equivalent isotropic displacement parameters (Å$^2$) for the crystal structures of $NdTa_7O_{19}$, $GdTa_7O_{19}$, and $ErTa_7O_{19}$.

| | Atom | *x* | *y* | *z* | $U_{iso}$*/$U_{eq}$ |
|---|---|---|---|---|---|
| $NdTa_7O_{19}$ | Ta1 | 0.3333 | 0.6667 | 0 | 0.00588(14) |
| | Ta2 | 0.63907(7) | 0.63951(6) | 0.15602(2) | 0.00639(6) |
| | Nd1 | 0.6667 | 0.3333 | 0 | 0.0069(2) |
| | O1 | 0.7512(10) | 1.0016(18) | 0.15375(15) | 0.0086(5) |
| | O2 | 0.5709(8) | 0.6222(7) | 0.0568(2) | 0.0072(6) |
| | O3 | 0.6246(10) | 0.5791(11) | 0.25 | 0.0068(9) |
| | O4 | 0.3333 | 0.6667 | 0.1681(4) | 0.0064(9) |
| | O5 | 0.6667 | 0.3333 | 0.1314(4) | 0.0072(11) |
| $GdTa_7O_{19}$ | Ta1 | 0.3333 | 0.6667 | 0 | 0.00573(14) |
| | Ta2 | 0.63905(7) | 0.63898(6) | 0.15581(2) | 0.00605(5) |
| | Gd1 | 0.6667 | 0.3333 | 0 | 0.0069(2) |
| | O1 | 0.7529(10) | 1.0037(18) | 0.15345(15) | 0.0080(5) |
| | O2 | 0.5718(7) | 0.6194(6) | 0.05606(19) | 0.0064(6) |
| | O3 | 0.6243(9) | 0.5791(10) | 0.25 | 0.0081(9) |
| | O4 | 0.3333 | 0.6667 | 0.1677(4) | 0.0073(9) |
| | O5 | 0.6667 | 0.3333 | 0.1310(4) | 0.0091(11) |
| $ErTa_7O_{19}$ | Ta1 | 0.3333 | 0.6667 | 0 | 0.0091(2) |
| | Ta2 | 0.63927(9) | 0.63829(8) | 0.15558(2) | 0.00901(8) |
| | Er1 | 0.6667 | 0.3333 | 0 | 0.0097(2) |
| | O1 | 0.7524(14) | 1.002(3) | 0.15316(19) | 0.0113(6) |
| | O2 | 0.5723(9) | 0.6156(8) | 0.0553(3) | 0.0103(9) |
| | O3 | 0.6226(11) | 0.5786(13) | 0.25 | 0.0100(12) |
| | O4 | 0.3333 | 0.6667 | 0.1663(5) | 0.0088(12) |
| | O5 | 0.6667 | 0.3333 | 0.1311(5) | 0.0106(14) |

**Table S4.** Selected bond distances (Å) and angles (°) for the crystal structure of $NdTa_7O_{19}$.

| Bond distance (Å) | | | |
|---|---|---|---|
| Nd—O2 | 2.435(4) | Ta2—O1 | 1.996(10) |
| Nd—O5 | 2.620(8) | Ta2—O2 | 2.014(4) |
| Ta1—O2 | 1.987(4) | Ta2—O3 | 1.9043(10) |
| Ta2—O1$^{vi}$ | 1.982(10) | Ta2—O4 | 2.0044(11) |
| Ta2—O1$^{iv}$ | 2.422(4) | Ta2—O5 | 2.054(2) |
| Angle (°) | | | |
| O2$^{v}$—Ta1—O2 | 166.2(2) | O3—Ta2—O5 | 94.0(3) |
| O2$^{ii}$—Ta1—O2 | 99.9(2) | O4—Ta2—O1$^{iv}$ | 64.8(2) |
| O2$^{iii}$—Ta1—O2 | 90.72(17) | O4—Ta2—O2 | 86.5(2) |
| O2$^{i}$—Ta1—O2 | 80.4(2) | O4—Ta2—O5 | 128.94(5) |
| O1—Ta2—O1$^{iv}$ | 137.8(3) | O5—Ta2—O1$^{iv}$ | 64.5(2) |
| O1$^{vi}$—Ta2—O1 | 85.0(2) | O2$^{i}$—Nd1—O2 | 63.53(19) |
| O1$^{vi}$—Ta2—O1$^{iv}$ | 137.1(3) | O2$^{ix}$—Nd1—O2 | 100.62(19) |
| O1—Ta2—O2 | 89.05(14) | O2$^{viii}$—Nd1—O2 | 155.57(19) |
| O1$^{vi}$—Ta2—O2 | 99.58(15) | O2$^{vii}$—Nd1—O2 | 100.11(11) |
| O1—Ta2—O4 | 73.68(12) | O2—Nd1—O5$^{i}$ | 117.72(9) |
| O1$^{vi}$—Ta2—O4 | 157.75(14) | O2—Nd1—O5 | 62.28(9) |
| O1$^{vi}$—Ta2—O5 | 73.30(13) | Ta2$^{x}$—O1—Ta2$^{iii}$ | 102.8(3) |
| O1—Ta2—O5 | 153.57(16) | Ta2—O1—Ta2iii | 102.1(3) |
| O2—Ta2—O1$^{iv}$ | 81.41(14) | Ta2$^{x}$—O1—Ta2 | 154.9(2) |
| O2—Ta2—O5 | 80.1(2) | Ta2$^{xi}$—O3—Ta2 | 159.5(3) |
| O3—Ta2—O1 | 101.53(19) | Ta2$^{iv}$—O4—Ta2 | 118.58(9) |
| O3—Ta2—O1$^{iv}$ | 84.01(18) | Ta2$^{vii}$—O5—Ta2 | 114.49(16) |
| O3—Ta2—O1$^{vi}$ | 91.41(18) | Ta2—O2—Nd1 | 111.94(16) |
| O3—Ta2—O2 | 165.4(2) | Ta2—O5—Nd1 | 103.8(2) |
| O3—Ta2—O4 | 86.9(3) | Ta1—O2—Nd1 | 108.06(17) |

Symmetry codes: (i) $-y+1, -x+1, -z$; (ii) $x, x-y+1, -z$; (iii) $-y+1, x-y+1, z$; (iv) $-x+y, -x+1, z$; (v) $-x+y, y, -z$; (vi) $-y+2, x-y+1, z$; (vii) $-y+1, x-y, z$; (viii) $x, x-y, -z$; (ix) $-x+y+1, y, -z$; (x) $-x+y+1, -x+2, z$; (xi) $x, y, -z+1/2$.

**Table S5.** Selected bond distances (Å) and angles (°) for the crystal structure of $GdTa_7O_{19}$.

| Bond distance (Å) | | | |
|---|---|---|---|
| Gd1—O2 (× 6) | 2.407(3) | Ta2—O1 | 2.007(9) |
| Gd1—O5 (× 2) | 2.607(8) | Ta2—O2 | 2.020(4) |
| Ta1—O2 (× 6) | 1.990(3) | Ta2—O3 | 1.9049(10) |
| Ta2—O1$^{viii}$ | 1.968(9) | Ta2—O4 | 2.0036(11) |
| Ta2—O1$^{vi}$ | 2.422(3) | Ta2—O5 | 2.050(2) |
| Angle (°) | | | |
| O2$^{iii}$—Ta1—O2 | 99.7(2) | O3—Ta2—O5 | 94.3(3) |
| O2$^{i}$—Ta1—O2 | 91.60(15) | O4—Ta2—O1$^{iv}$ | 65.1(2) |
| O2$^{v}$—Ta1—O2 | 165.3(2) | O4—Ta2—O2 | 86.8(2) |
| O2—Ta1—O2$^{ii}$ | 78.9(2) | O4—Ta2—O5 | 129.00(5) |
| O1$^{vi}$—Ta2—O1$^{iv}$ | 137.0(3) | O5—Ta2—O1$^{iv}$ | 64.4(2) |
| O1$^{vi}$—Ta2—O1 | 84.67(18) | O2$^{ix}$—Gd1—O2 | 100.22(11) |
| O1—Ta2—O1$^{iv}$ | 138.2(3) | O2$^{ii}$—Gd1—O2 | 63.36(17) |
| O1$^{vi}$—Ta2—O2 | 99.24(14) | O2$^{viii}$—Gd1—O2 | 155.53(18) |
| O1—Ta2—O2 | 89.45(13) | O2$^{vii}$—Gd1—O2 | 100.56(18) |
| O1$^{vi}$—Ta2—O4 | 157.58(12) | O2—Gd1—O5$^{ii}$ | 117.63(9) |
| O4—Ta2—O1 | 73.76(11) | O2—Gd1—O5 | 62.37(8) |
| O1$^{vi}$—Ta2—O5 | 73.42(11) | Ta2$^{xi}$—O1—Ta2$^{i}$ | 102.9(3) |
| O1—Ta2—O5 | 153.33(16) | Ta2$^{xi}$—O1—Ta2 | 155.15(17) |
| O2—Ta2—O1$^{iv}$ | 81.29(13) | Ta2—O1—Ta2$^{i}$ | 101.7(3) |
| O2—Ta2—O5 | 79.5(2) | Ta2—O3—Ta2$^{xi}$ | 159.7(3) |
| O3—Ta2—O1 | 101.46(18) | Ta2$^{i}$—O4—Ta2 | 118.63(9) |
| O3—Ta2—O1$^{iv}$ | 84.05(18) | Ta2$^{x}$—O5—Ta2 | 114.37(17) |
| O3—Ta2—O1$^{vi}$ | 91.57(17) | Ta2—O5—Gd1 | 104.0(2) |
| O3—Ta2—O2 | 165.3(2) | Ta2—O2—Gd1 | 112.37(14) |
| O3—Ta2—O4 | 86.9(3) | Ta1—O2—Gd1 | 108.87(16) |

Symmetry codes: (i) $-y+1, x-y+1, z$; (ii) $-y+1, -x+1, -z$; (iii) $x, x-y+1, -z$; (iv) $-x+y, -x+1, z$; (v) $-x+y, y, -z$; (vi) $-y+2, x-y+1, z$; (vii) $-x+y+1, y, -z$; (viii) $x, x-y, -z$; (ix) $-x+y+1, -x+1, z$; (x) $-y+1, x-y, z$; (x) $-x+y+1, -x+2, z$; (xi) $x, y, -z+1/2$.

**Table S6.** Selected bond distances (Å) and angles (°) for the crystal structure of $ErTa_7O_{19}$.

| Bond distance (Å) | | | |
|---|---|---|---|
| Er1—O2 | 2.373(4) | Ta2—O1 | 1.997(13) |
| Er1—O5 | 2.603(10) | Ta2—O2 | 2.024(5) |
| Ta1—O2 | 1.993(5) | Ta2—O3 | 1.9042(12) |
| Ta2—O1$^{viii}$ | 1.972(13) | Ta2—O4 | 2.0011(13) |
| Ta2—O1$^{iv}$ | 2.419(4) | Ta2—O5 | 2.039(3) |
| Angle (°) | | | |
| O2—Ta1—O2$^{ii}$ | 92.5(2) | O3—Ta2—O5 | 94.4(3) |
| O2—Ta1—O2$^{i}$ | 99.9(3) | O4—Ta2—O1$^{ii}$ | 64.8(3) |
| O2—Ta1—O2$^{iii}$ | 77.3(3) | O4—Ta2—O2 | 86.6(3) |
| O2—Ta1—O2$^{v}$ | 164.2(3) | O4—Ta2—O5 | 128.94(7) |
| O1—Ta2—O1$^{ii}$ | 137.7(4) | O5—Ta2—O1$^{ii}$ | 64.7(3) |
| O1$^{vi}$—Ta2—O1 | 84.4(2) | O2$^{viii}$—Er1—O2 | 100.7(2) |
| O1$^{vi}$—Ta2—O1$^{ii}$ | 137.7(4) | O2$^{vii}$—Er1—O2 | 155.4(2) |
| O1$^{vi}$—Ta2—O2 | 99.00(18) | O2—Er1—O2$^{iii}$ | 63.2(2) |
| O1—Ta2—O2 | 89.93(16) | O2$^{ix}$—Er1—O2 | 100.27(15) |
| O1$^{vi}$—Ta2—O4 | 157.35(16) | O2$^{iii}$—Er1—O5 | 117.59(12) |
| O1—Ta2—O4 | 73.58(15) | O2—Er1—O5 | 62.41(12) |
| O1—Ta2—O5 | 153.5(2) | Ta2$^{x}$—O1—Ta2 | 155.4(2) |
| O1$^{vi}$—Ta2—O5 | 73.71(15) | Ta2$^{x}$—O1—Ta2$^{iv}$ | 102.2(4) |
| O2—Ta2—O1$^{ii}$ | 81.00(17) | Ta2—O1—Ta2$^{iv}$ | 102.2(4) |
| O2—Ta2—O5 | 79.1(3) | Ta2—O3—Ta2$^{xi}$ | 160.0(4) |
| O3—Ta2—O1$^{ii}$ | 84.0(2) | Ta2$^{ii}$—O4—Ta2 | 118.88(10) |
| O3—Ta2—O1 | 101.4(2) | Ta2$^{ix}$—O5—Ta2 | 114.5(2) |
| O3—Ta2—O1$^{vi}$ | 92.0(2) | Ta2—O2—Er1 | 112.90(19) |
| O3—Ta2—O2 | 165.0(3) | Ta2—O5—Er1 | 103.8(3) |
| O3—Ta2—O4 | 87.2(3) | Ta1—O2—Er1 | 109.8(2) |

Symmetry codes: (i) $x$, $x-y+1$, $-z$; (ii) $-x+y$, $-x+1$, $z$; (iii) $-y+1$, $-x+1$, $-z$; (iv) $-y+1$, $x-y+1$, $z$; (v) $-x+y$, $y$, $-z$; (vi) $-y+2$, $x-y+1$, $z$; (vii) $x$, $x-y$, $-z$; (viii) $-x+y+1$, $y$, $-z$; (ix) $-y+1$, $x-y$, $z$; (x) $-x+y+1$, $-x+2$, $z$; (xi) $x$, $y$, $-z+1/2$.